# Algae-Filler Artificial Timber with an Ultralow Binder Content

Haozhe Yi,[1] Kiwon Oh,[2] Rui Kou,[1] Yu Qiao,[1,2,*]

[1] *Department of Structural Engineering, University of California – San Diego, La Jolla, CA 92093-0085, U.S.A.*

[2] *Program of Materials Science and Engineering, University of California – San Diego, La Jolla, CA 92093, U.S.A.*

[*] *Corresponding author. Phone: +1-858-534-3388; Email: yqiao@ucsd.edu*

**Abstract:** Algae cultivation is an active area of study for carbon sequestration, while the large amount of produced algae must be upcycled. In the current study, we fabricated artificial timber based on algae filler, with only 2~4% epoxy binder. The flexural strength could be comparable with those of softwoods. The binder was efficiently dispersed in the algae phase through diluent-aided compaction self-assembly. The important processing parameters included the binder content, the filler morphology, the compaction pressure, the diluent ratio, and the curing condition. This research not only is critical to carbon sequestration, but also helps reduce the consumption of conventional construction materials.

**Graphic TOC**

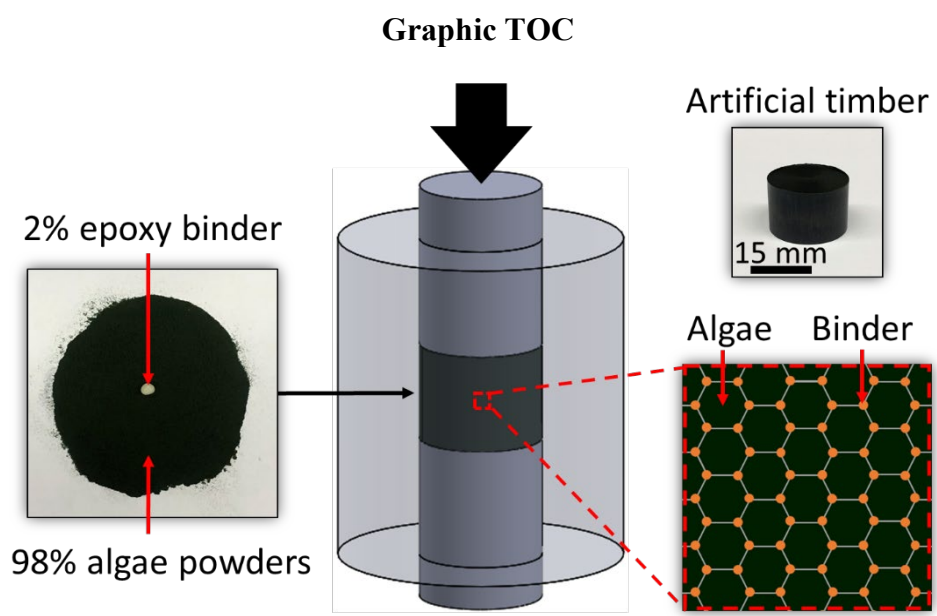

# 1. Introduction

A promising method of carbon sequestration is to cultivate algae [1]. Algae is low maintenance and fast growing [2,3]. Compared to trees, algae can absorb carbon dioxide ($CO_2$) more efficiently by two orders of magnitude [3–5]. If equipped with bioreactor-enabled algae farms, a large city may have a negative overall carbon emission rate [6].

The major challenges to this concept are related to the large scale of algae production. To achieve a nontrivial impact on the $CO_2$ amount in the atmosphere, each year billions of tons of algae must be produced [7]. They should not decompose back to $CO_2$. Permanent storage can be costly. It involves proper dehydration, transportation, and placement. One attractive approach is to use algae to help meet the increasingly high demand of food security [8]. Because algae is not traditionally a main food source, it must be processed to separate proteins and carbonhydrates, for which basic research is being carried out [9,10]. Another area of study is to utilize algae as fuel and biomass [11,12]. The main issues include the relatively high cost and the difficulties in scaling up. The cellulose content in cell walls of algae is in the range from 40% to 70% [13], offering an opportunity to fabricate structural parts. Dehydrated algae has been converted to polymers and foams [14–16].

Among all the potential applications, construction materials form a sufficiently large market. Every year, ~10 billion tons concrete are consumed worldwide [17], responsible to 12~15% of total industrial energy use [18] and 5~8% of total human-related $CO_2$ emission [19]. In addition, a few billion $m^3$ wood is used annually [20]. If algae-based materials can replace a portion of concrete and timber, it not only economically helps algae cultivation, but also reduces the energy use and carbon emission from the conventional construction industry.

Dehydrated algae often exists as non-cohesive powders. To apply it in engineering structures, particulate composites need to be processed. A regular particulate composite contains 70~80% filler particles and 20~30% binder [21]. The fillers are usually sand, wood chips, carbon black, etc. [22]. The binder can be epoxy, unsaturated polyester resin, vinyl ester, phenolics, or a thermoplastic [23]. These binders are relatively expensive, and their production emits $CO_2$ [23–25]. If we use 30% binder in an algae-filler composite, the cost-performance balance and the overall environmental benefit would be unsatisfactory. It is desirable to largely decrease the binder content to below 5%.

With such an ultralow binder content, ordinary composite processing techniques are no longer relevant. The filler-binder wetting would be poor, and the final defect density would be large. To solve these problems, we recently developed the compaction self-assembly (CSA) technology [26–33]. In CSA, filler and binder are first premixed, and then compacted under a relatively high pressure. The binder content can be only ~4%, and the maximum compaction pressure is generally 30~100 MPa. The compaction pressure not only squeezes the binder droplets and densify the interstitial gaps, but also deforms and rotates the filler particles. More critically, as the filler particles are close-packed, at the direct contact points, a large capillary pressure would be built up, driving the binder to these most important microstructural sites. As the binder is self-assembled into binder micro-agglomerations (BMA), the load-carrying capacity is optimized.

In the current study, we investigate how to fabricate ultralow-binder-content algae-filler composites. The produced material will be referred to as artificial timber. If successful, the research will have profound impacts on the development of green construction materials as well as the study on algae-related $CO_2$ sequestration.

## 2. Experimental procedure

We investigated two types of algae, denoted by FP and SW, respectively. Algae FP was obtained from Good Natured in powder form (Product No. 857307002257), with the average particle size of 10~15 μm. Algae SW was prepared from zostera marina harvested at the La Jolla Shore, La Jolla, California. The SW sample was thoroughly rinsed and dried in a mechanical convection oven (Jeio Tech OF-12G-120) at 80 °C for 24 hours, chopped by a razor blade into cm-sized pieces, and ground in a MTI MSK-SFM-14 roller mill at 50 rpm for 2 h. The particle size of the milled SW was around 50 μm to 0.5 mm.

Figure 1 illustrates the material processing procedure. Epoxy resin (Hexion, Epon 828) was employed as the binder, with the curing agent being Hexion Epikure-3115 polyamide. For each part of epoxy, 1.2 parts curing agent and a certain amount of isopropyl alcohol (IPA) were added, and manually mixed in a glass vial at room temperature (~22 °C) for 20 min. The IPA to epoxy ratio, $\alpha_I = m_{IPA}/m_{eh}$, ranged from 2.5 to 6, where $m_{IPA}$ and $m_{eh}$ are the IPA mass and the epoxy-hardener mixture mass, respectively. Powders of algae filler were blended with the IPA-diluted epoxy in a Thinky ARE-310 centrifugal mixer at 2000 rpm for 3.5 min. The binder content, $\alpha_b = m_{eh}/(m_{eh} + m_a)$, was in the range from 2% to 10%, where $m_a$ is the algae mass. The material was transferred into a steel cylindrical mold. Two pistons were inserted into the mold from both ends. The height, the outer diameter, and the inner diameter of the mold were 50.7 mm, 44.5 mm, and 19.1 mm, respectively. The height and the diameter of the piston were 25.4 mm and 19.1 mm, respectively. Compaction self-assembly was carried out by compressing the algae-binder mixture in a universal testing machine (Instron 5582) at the loading rate of 15 mm/min, until the desired peak pressure ($P_c$) was reached. The loading plate held the peak pressure for 1 min. The peak pressure ranged from 10 MPa to 350 MPa. The two pistons were then fixed by a C-clamp (McMaster-Carr, 5133A19). The mold was placed in the convection oven, and the binder was hardened at 100 °C for 1 hour.

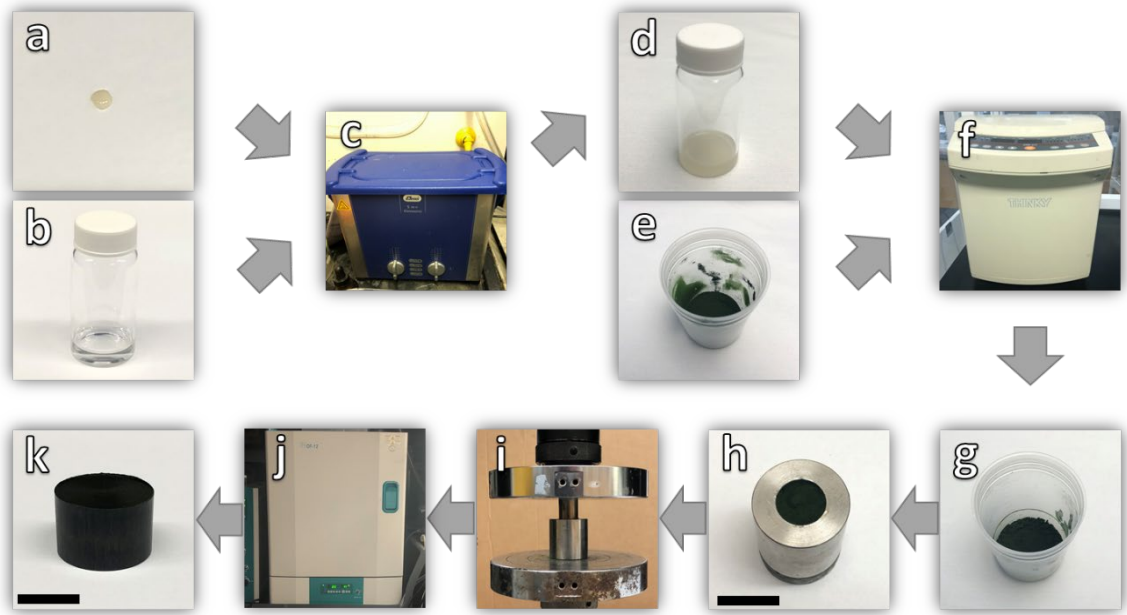

**Figure 1.** The processing procedure: (a) mixed epoxy resin and hardener, (b) isopropyl alcohol (IPA), (c) the sonication bath, (d) the IPA-diluted epoxy, (e) the milled algae powders, (f) the paste mixer, (g) the algae-binder mixture, (h) the steel mold (scale bar: 30 mm), (i) the setup of compaction self-assembly (CSA), (j) curing in a convection oven, and (k) an artificial timber sample (scale bar: 12.5 mm).

After curing, the mold was air-cooled for 1 hour, and the artificial timber sample was pushed out by a press. Flexural testing specimens were sectioned by a high-speed diamond saw (MTI, SYJ-40-LD), with the length, depth ($d$), and width ($w$) being 18 mm, 5 mm, and 5mm, respectively. The surfaces of the specimen were flattened by 400-grit sandpapers. The flexural strength was measured through three-point bending in the Instron testing machine. The specimen was supported at both ends by two 2.54-mm-diameter 19-mm-long stainless steel pins. Another steel pin rested at the top of the beam specimen at the middle, and was compressed downwards at the rate of 6 mm/min, until the specimen failed. The peak loading, $F_f$, was recorded. The flexural strength was calculated as $R = \frac{3}{2}\frac{F_f L}{wd^2}$, with $L = 16$ mm being the distance between the two supporting pins. At least three specimens were tested for each condition. Scanning electron microscope (SEM) images were taken at the fracture surfaces.

## 3. Result and discussions

Figure 2(a) shows the measured flexural strength of artificial timber samples. The binder content varies from 2% to 10%. Remarkably, with only 2% binder, the strength of FP-filler samples can reach ~20 MPa, stronger than typical portland cement by 5~10 times and better than most steel-reinforced concrete [34]. As shown in Fig.1(k), the compacted material is blackish, having a smooth surface. As the binder content ($\alpha_b$) increases from 2% to 8%, $R$ is improved somewhat linearly. When $\alpha_b = 8\%$, $R$ is ~27 MPa. When $\alpha_I$ further rises to 10%, the strength remains similar; that is, the binder distribution has saturated. This is consistent with our previous result [32] that as the filler is densified, the minimum gap volume is ~8%. Excess binder would be squeezed out of the materials system, and does not contribute to the final structural integrity.

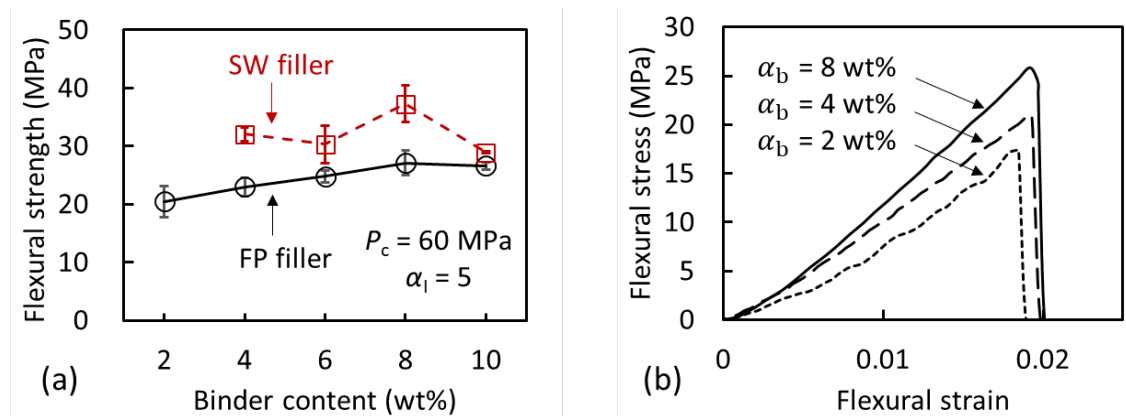

**Figure 2.** (a) The flexural strength ($R$) as a function of the binder content ($\alpha_b$). (b) Typical stress-strain curves of FP-filler samples; the IPA to epoxy mass ratio ($\alpha_I$) is 5, and the compaction pressure ($P_c$) is 60 MPa.

The stress-strain curves in Fig.2(b) show that as the binder content increases, both the strength and the stiffness become larger. The increase in strength is more pronounced, so that the failure strain rises with $\alpha_b$. Because the samples are formed by small powders, when the final cracking begins, no long fibers could bridge the crack flanks and consequently, the fracture process is rapid.

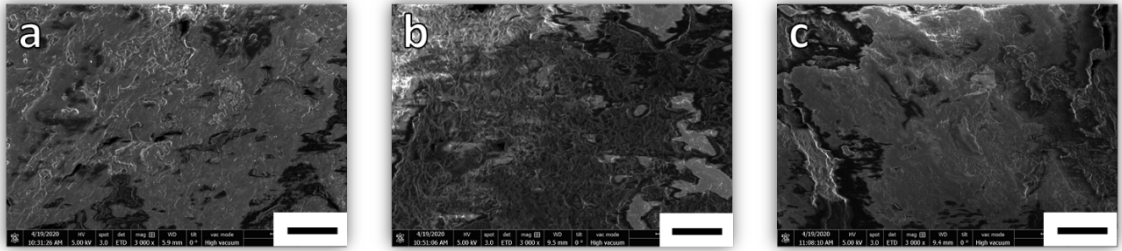

**Figure 3.** Typical SEM images of FP-filler samples, with (a) $\alpha_b = 2$ wt%, (b) $\alpha_b = 6$ wt%, and (c) $\alpha_b = 10$ wt% (scale bar: 10 μm).

The relatively high strength should be attributed to the efficient BMA formation, as well as the filler densification. As shown in Fig.3(a), when $\alpha_b = 2\%$, the morphology of the material is already quite homogeneous. Few large-sized defects can be observed. Since algae is deformable upon compaction, during CSA the interstitial sites are largely reduced, and the premixed binder droplets necessarily aggregate into micro-agglomerations at the angular edges of adjacent filler powders. The binder motion is promoted by the capillary pressure, as the CSA pressure greatly decreases the characteristic size of the microstructural channels. As a result, the small amount of binder is dispersed as an efficient load-transferring network, connecting the algae powders into a solid. Such a dense microstructure is maintained in the curing process, as the steel pistons are clamped. If the pistons were free and the material were allowed to expand, the strength of the cured sample would be lowered by 3~4 times. Clearly, before hardening, the binder cannot lock the BMA structure. The thermal mismatch among various components would interrupt the close-packed filler-binder mixture.

When more binder is used, as indicated by Fig.3(b), the homogeneity is improved and the defects formation is suppressed, leading to a higher strength. As the effective porosity is reduced, the modulus of elasticity is larger, causing the observed stiffening effect (Fig.2b). When the binder content is 10%, the material is similar to a conventional particulate composite, wherein the binder forms a continuous matrix and the algae powders are fully embedded. This critical value of binder content is 2~3 times less than that of regular composites, thanks to the compaction pressure that deforms, rotates, and compresses the algae phase.

According to Fig.2(a), the SW-filler samples generally exhibit a higher strength than the FP-filler samples. Because the SW powder size is considerably larger that of FP, SW tends to have a larger aspect ratio, somewhat similar to microfibers. Thus, in addition to the bonding through BMA, the SW particles may be entangled, further enhancing the structural properties.

With only 4% binder, the flexural strength is ~33 MPa, comparable with softwoods [35]. Yet, if the binder content is reduced to 2%, the material would be quite weak and the flexural strength measurement becomes difficult. It is interesting that the flexural strength is not sensitive to the binder content in the range of $\alpha_b$ from 4% to 10%. As a considerable portion of strength comes from the internal friction and entanglement of algae, the role of binder is still important, but different from in FP-filler samples. The binder provides distributed locking sites that hold the algae together. As long as the MBA network is relatively complete, the material strength would be optimized.

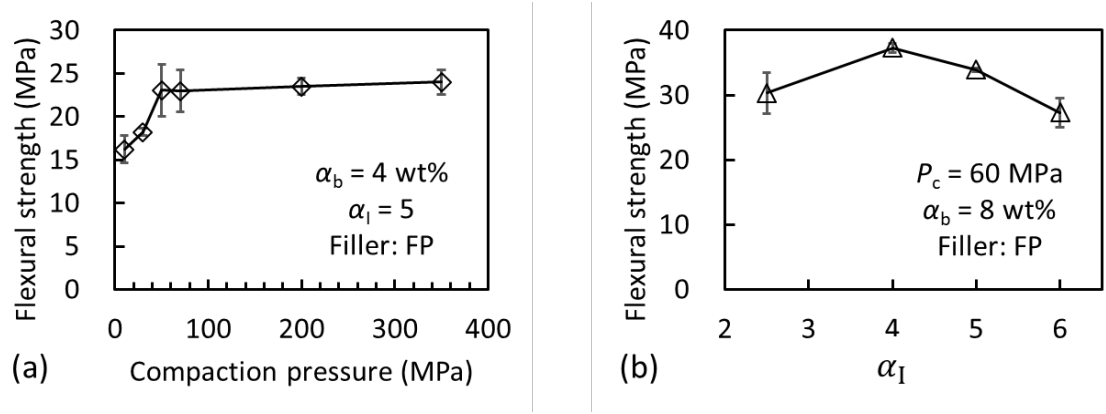

**Figure 4.** The flexural strength ($R$) as a function of (a) the compaction pressure ($P_c$) and (b) the IPA to binder mass ratio ($\alpha_I$).

In addition to the binder content and the filler type, another important processing parameter is the compaction pressure, $P_c$. As suggested by Fig.4(a), in general, a higher compaction pressure is beneficial, especially when it is less than 60 MPa. When $P_c = 10$ MPa, the flexural strength is ~16 MPa. When $P_c$ rises to 60 MPa, the strength increases nearly linearly to ~22 MPa by ~40%. Beyond 60 MPa, the effect of $P_c$ is secondary. When $P_c$ is very high ~350 MPa, $R$ is ~24 MPa, only marginally larger than the strength of $P_c = 60$ MPa. It is clear that the densification effect saturates at ~60 MPa, at which the algae powders have collapsed and the powder rotation and sliding have nearly completed. A compaction pressure around 30~60 MPa is on the same scale as the pressure of compression molding [36,37]. To achieve a high strength, the optimum $P_c$ is ~60 MPa.

The IPA to epoxy-hardener ratio ($\alpha_I$) affects the material structure by changing the rheological properties of the binder. As demonstrated in Fig.4(b), when $\alpha_I$ is less than 4, using more IPA helps enhance the strength, since the binder viscosity is reduced and the binder dispersion is more widespread. When $α_I$ exceeds 4, further increasing it would have a detrimental effect, probably because of the increase in defect density, associated with the volume left by evaporated IPA. IPA may also interrupt the epoxy ring opening, as the functional groups of polyamide are obstructed. The optimum $α_I$ is ~4, at which the flexural strength reaches ~37 MPa, similar to those of aspen, basswood, sassafras, etc. [35].

**4. Conclusions**

We produced artificial timber using algae as the filler and epoxy as the binder. The binder content could be only 2~4 wt%, and the flexural strength could be comparable with those of softwoods. The material was quite homogeneous, with a low defect density. The critical processing operation was the compaction self-assembly. Upon a sufficiently high compression pressure, algae powders were deformed, rotated, and densified, and the binder phase was relatively uniformly dispersed. A large filler powder size and aspect ratio helped improve the structural integrity. The optimum compaction pressure was ~60 MPa; further increasing it would not lead to much enhancement in strength. The binder viscosity was critical. It should be controlled by a diluent, e.g., IPA. During curing, the material should to be confined, so that the compacted microstructure could remain. This study may provide an important method to upcycle algae, critical not only to the study of algae cultivation and carbon sequestration, but also to the development of next-generation green construction materials.

**Acknowledgement**

This work was supported by ARPA-E under Grant No. DE-AR0001144. The materials characterization was performed in part at the San Diego Nanotechnology Infrastructure (SDNI) of the University of California – San Diego, a member of the National Nanotechnology Coordinated Infrastructure, which is supported by the National Science Foundation (Grant No. ECCS-1542148).

**Reference**

[1]   K. Soratana, A.E. Landis, Evaluating industrial symbiosis and algae cultivation from a

life cycle perspective, Bioresour. Technol. 102 (2011) 6892–6901.

[2] H. Blaas, C. Kroeze, Possible future effects of large-scale algae cultivation for biofuels on coastal eutrophication in Europe, Sci. Total Environ. 496 (2014) 45–53.

[3] F. Bux, Y. Chisti, Algae biotechnology: products and processes, Springer, 2016.

[4] P. Talbot, M.P. Gortares, R.W. Lencki, J. De la Noüe, Absorption of CO2 in algal mass culture systems: a different characterization approach, Biotechnol. Bioeng. 37 (1991) 834–842.

[5] Y. Lee, S.J. Pirt, CO2 absorption rate in an algal culture: effect of pH, J. Chem. Technol. Biotechnol. Biotechnol. 34 (1984) 28–32.

[6] D. Moreira, J.C.M. Pires, Atmospheric CO2 capture by algae: negative carbon dioxide emission path, Bioresour. Technol. 215 (2016) 371–379.

[7] Z. Chi, F. Elloy, Y. Xie, Y. Hu, S. Chen, Selection of microalgae and cyanobacteria strains for bicarbonate-based integrated carbon capture and algae production system, Appl. Biochem. Biotechnol. 172 (2014) 447–457.

[8] J.E. Duffy, E.A. Canuel, W. Adey, J.P. Swaddle, Biofuels: algae, Science. 326 (2009) 1345.

[9] M.L. Wells, P. Potin, J.S. Craigie, J.A. Raven, S.S. Merchant, K.E. Helliwell, A.G. Smith, M.E. Camire, S.H. Brawley, Algae as nutritional and functional food sources: revisiting our understanding, J. Appl. Phycol. 29 (2017) 949–982.

[10] I. Wijesekara, S. Kim, Application of marine algae derived nutraceuticals in the food industry, Mar. Algae Extr. Process. Prod. Appl. 35 (2015) 627–638.

[11] M. Dębowski, M. Zieliński, A. Grala, M. Dudek, Algae biomass as an alternative substrate in biogas production technologies, Renew. Sustain. Energy Rev. 27 (2013) 596–604.

[12] A.B.M.S. Hossain, A. Salleh, A.N. Boyce, P. Chowdhury, M. Naqiuddin, Biodiesel fuel production from algae as renewable energy, Am. J. Biochem. Biotechnol. 4 (2008) 250–254.

[13] A. Mihranyan, Cellulose from cladophorales green algae: From environmental problem to high-tech composite materials, J. Appl. Polym. Sci. 119 (2011) 2449–2460.

[14] A.J. Rubin, E.A. Cassel, O. Henderson, J.D. Johnson, J.C. Lamb III, Microflotation: New low gas-flow rate foam separation technique for bacteria and algae, Biotechnol. Bioeng. 8 (1966) 135–151.

[15] M.A. Ragan, J.S. Craigie, Physodes and the phenolic compounds of brown algae. Isolation and characterization of phloroglucinol polymers from Fucus vesiculosus (L.),

Can. J. Biochem. 54 (1976) 66–73.

[16] P. Fasahati, J.J. Liu, Economic, energy, and environmental impacts of alcohol dehydration technology on biofuel production from brown algae, Energy. 93 (2015) 2321–2336.

[17] C. Meyer, The greening of the concrete industry, Cem. Concr. Compos. 31 (2009) 601–605.

[18] N.A. Madlool, R. Saidur, M.S. Hossain, N.A. Rahim, A critical review on energy use and savings in the cement industries, Renew. Sustain. Energy Rev. 15 (2011) 2042–2060.

[19] E. Benhelal, G. Zahedi, E. Shamsaei, A. Bahadori, Global strategies and potentials to curb CO2 emissions in cement industry, J. Clean. Prod. 51 (2013) 142–161.

[20] A.H. Buchanan, S.B. Levine, Wood-based building materials and atmospheric carbon emissions, Environ. Sci. Policy. 2 (1999) 427–437.

[21] C.A. Harper, Handbook of plastics, elastomers, and composites, McGraw-Hill New York, 2002.

[22] N.C. Consoli, J.P. Montardo, M. Donato, P.D. Prietto, Effect of material properties on the behaviour of sand—cement—fibre composites, Proc. Inst. Civ. Eng. Improv. 8 (2004) 77–90.

[23] J.-P. Pascault, H. Sautereau, J. Verdu, R.J.J. Williams, Thermosetting polymers, CRC press, 2002.

[24] J.-P. Pascault, R.J.J. Williams, Epoxy polymers: new materials and innovations, John Wiley & Sons, 2009.

[25] M. Biron, Thermosets and composites: material selection, applications, manufacturing and cost analysis, Elsevier, 2013.

[26] K. Oh, T. Chen, A. Gasser, R. Kou, Y. Qiao, Compaction self-assembly of ultralow-binder-content particulate composites, Compos. Part B Eng. 175 (2019) 107144.

[27] B.J. Chow, T. Chen, Y. Zhong, Y. Qiao, Direct formation of structural components using a martian soil simulant, Sci. Rep. 7 (2017) 1–8.

[28] B.J. Chow, T. Chen, Y. Zhong, M. Wang, Y. Qiao, Compaction of montmorillonite in ultra-dry state, Adv. Sp. Res. 60 (2017) 1443–1452.

[29] R. Kou, Y. Zhong, J. Kim, Q. Wang, M. Wang, R. Chen, Y. Qiao, Elevating low-emissivity film for lower thermal transmittance, Energy Build. 193 (2019) 69–77.

[30] Y. Zhong, R. Kou, M. Wang, Y. Qiao, Synthesis of centimeter-scale monolithic SiC nanofoams and pore size effect on mechanical properties, J. Eur. Ceram. Soc. 39 (2019)

2566–2573.

[31] H. Su, Y. Hong, T. Chen, R. Kou, M. Wang, Y. Zhong, Y. Qiao, Fatigue behavior of inorganic-organic hybrid "lunar cement," Sci. Rep. 9 (2019) 1–8.

[32] T. Chen, B.J. Chow, M. Wang, Y. Zhong, Y. Qiao, High-pressure densification of composite lunar cement, J. Mater. Civ. Eng. 29 (2017) 6017013.

[33] T. Chen, B.J. Chow, Y. Zhong, M. Wang, R. Kou, Y. Qiao, Formation of polymer micro-agglomerations in ultralow-binder-content composite based on lunar soil simulant, Adv. Sp. Res. 61 (2018) 830–836.

[34] K. Leet, D. Bernal, Reinforced concrete design, McGraw-Hill New York, 1982.

[35] D.W. Green, J.E. Winandy, D.E. Kretschmann, Mechanical properties of wood, Wood Handb. Wood as an Eng. Mater. Madison, WI USDA For. Serv. For. Prod. Lab. 1999. Gen. Tech. Rep. FPL; GTR-113 Pages 4.1-4.45. 113 (1999).

[36] R.A. Tatara, Compression molding, in: Appl. Plast. Eng. Handb., Elsevier, 2017: pp. 291–320.

[37] H. Suherman, A.B. Sulong, J. Sahari, Effect of the compression molding parameters on the in-plane and through-plane conductivity of carbon nanotubes/graphite/epoxy nanocomposites as bipolar plate material for a polymer electrolyte membrane fuel cell, Ceram. Int. 39 (2013) 1277–1284.